# Hysteretic behaviour of metal connectors for hybrid (high- and low-grade mixed species) cross laminated timber

Mahdavifar, V.[1], Barbosa, A.R.[2], Sinha, A.[3], Muszynski, L.[4], Gupta, R.[5]

**ABSTRACT:** Cross-laminated timber (CLT) is a prefabricated solid engineered wood product made of at least three orthogonally bonded layers of solid-sawn lumber that are laminated by gluing longitudinal and transverse layers with structural adhesives to form a solid panel. Previous studies have shown that the CLT buildings can perform well in seismic loading and are recognized as the essential role of connector performance in structural design, modelling, and analysis of CLT buildings. When CLT is composed of high-grade/high-density layers for the outer lamellas and low-grade/low-density for the core of the panels, the CLT panels are herein designated as hybrid CLT panels as opposed to conventional CLT panels that are built using one lumber type for both outer and core lamellas.

This paper presents results of a testing program developed to estimate the cyclic performance of CLT connectors applied on hybrid CLT layups. Two connectors are selected, which can be used in wall-to-floor connections. These are readily available in the North American market. Characterization of the performance of connectors is done in two perpendicular directions under a modified CUREE cyclic loading protocol. Depending on the mode of failure, in some cases, testing results indicate that when the nails or screws penetrate the low-grade/low-density core lumber, a statistically significant difference is obtained between hybrid and conventional layups. However, in other cases, due to damage in the face layer or in the connection, force-displacement results for conventional and hybrid CLT layups were not statistically significant.

## 1 INTRODUCTION

Cross-laminated timber (CLT) is a prefabricated solid engineered wood product made of at least three orthogonally bonded layers of solid-sawn lumber that are laminated by gluing of longitudinal and transverse layers, with structural adhesives to form a solid panel intended for roof, floor, or wall applications. The CLT manufacturing process and the construction technology based on this product has been used in Europe for over 25 years and recently started to emerge into the North American market.

CLT building structures have been shown to perform well in seismic loading [1]. The structural performance of the CLT structures mostly depends on the performance of their components: the panels and the connections. The major form for CLT construction is platform construction in which the loads are transferred through the floor panels to the walls of the corresponding story, then to lower stories, all the way down to the foundation. For the gravity loads, load transfer between the panels is developed through the contact surfaces between the panels. Under lateral loads, the CLT shear walls (panels and connectors) carry most of the loads. The lateral resistance of CLT shear walls is significantly affected by the type, arrangement, and location of the connectors, as well as the friction between the panel and the floor or foundation.

Four major displacement mechanisms, as shown in Figure 1, are possible in a CLT wall subjected to lateral in-plane loading: (1) rocking, (2) sliding, (3) flexure deformation, and (4) shear deformation. Since the CLT panels have relatively high in-plane stiffness in comparison to connectors, the displacements due to flexure and shear deformation are small. Thus, displacements in CLT walls are typically governed by sliding and rocking mechanisms. The rocking and sliding will cause two distinct perpendicular forces on the connections: withdrawal and shear. Angle bracket connections (A1 through A4 in Figure 1) provide resistance to the shear force but little withdrawal resistance. Due to the low value of withdrawal resistance of the angle brackets, uplift forces due to rocking are usually resisted by hold-downs placed at ends of walls (see H1 and H2 in Figure 1).

In the presence of the mechanisms shown in Figure 1, capacity design principles are recommended for design of CLT wall panels [2]. In this method of design, desirable modes of failure include connection ductile failure, wood crushing, or a combination thereof prior to onset of undesirable brittle wood failure modes. Since

1. Vahid Mahdavifar, Oregon State University, vahid.mahdavifar@oregonstate.edu
2. Andre Barbosa, Oregon State University, Andre.barbosa@oregonstate.edu
3. Arijit Sinha, Oregon State University, arijit.sinha@oregonstate.edu
4. Lech Muszyński, Oregon State University, lech.muszynski@oregonstate.edu
5. Rakesh Gupta, Oregon State University, rakesh.gupta@oregonstate.edu

the CLT panels have high in-plane stiffness relative to the connectors, a major portion of the energy dissipation capacity is provided by the metal connections, while the CLT wall panels remain in the elastic domain. The same capacity design principles should be applied to the whole CLT structure.

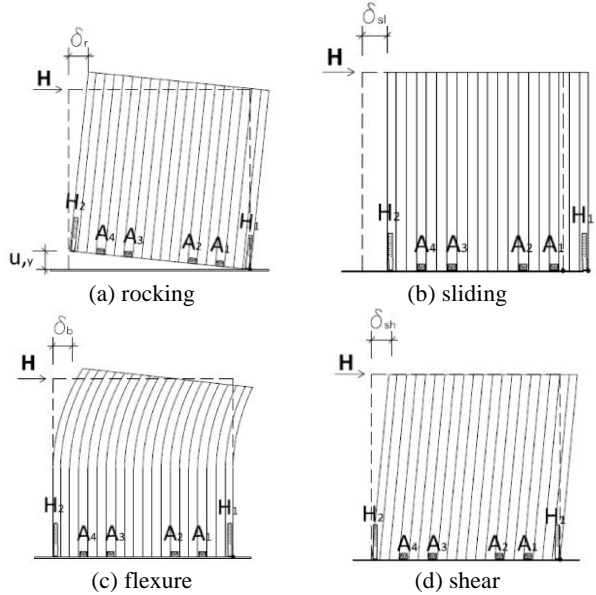

*Figure 1: Displacement mechanisms of wall panel subjected to lateral in-plane load due to (a) rocking, (b) sliding, (c) flexure, and (d) shear* **[3]**.

As CLT construction emerges into new markets such as mid-rise buildings, there is a need for optimizing the use of timber. In the North America, nearly all CLT structures being designed and constructed use CLT products that are manufactured following 2012 ANSI/APA Standard for Performance Rated Cross Laminated Timber (ANSI PRG-320) [4]. However, it is expected that as the market will continue to mature with time, more diversified grades and special CLT products will make its way into the market. One special CLT product could be a *hybrid CLT* panel. Throughout this paper, a hybrid CLT panel refers to a panel with layups arranged from high- and low-grade lamellas, but the layups could also be arranged from high- and low-density materials and using different species in different layers. This is a new concept aiming towards improving the economic efficiency and sustainability of the CLT industry as well as optimizing raw material resources. Utilization of low-value lumber in value-added engineered products such as hybrid CLT should reduce the pressure on the high end structural lumber supply and will also provide a substantial outlet for harvested lower value lumber.

In a hybrid CLT panel, the outer layers will typically comprise of high grade or higher density lumber, while the inner layers can be made up of low grade/low-density species.

The connections used in CLT structures are specialty connections that use different mechanical fasteners, currently not listed in the National Design Specifications for Wood Construction (NDS) [5]. In addition, connection performance depends mainly on the bending strength of the fastener and the density of the wood substrate to which they embed. The variation in density profile in a hybrid CLT panel along the panel thickness may cause fasteners to be embedded in layers of varying density. As a result, the simple design equations provided in the NDS based on the European Yield Model (EYM) [6] are not applicable to predict the fastener system performance.

In summary, understanding the performance of the connections is crucial in design of CLT buildings, as the connections provide ductility, strength, and stiffness. In this paper, partial preliminary results of a larger research program currently underway at Oregon State University (OSU) is presented. The larger research program is investigating hybrid layups with under-utilized or under-valued North American species and low-grade timber harvested in the region. The results presented provide the characterization of the performance of connectors in two perpendicular directions, under a modified CUREE cyclic loading protocol.

## 2 PREVIOUS RESEARCH

Several research projects have been recently completed to better understand behaviour of CLT connections, both in Europe and North America. With respect to structural laboratory testing, Ceccotti et al. [1] tested the lateral resistance of CLT walls. Different configurations of wall panels were investigated by means of monotonic and cyclic displacement. It was observed that the opening layout and design of the joints influenced the overall behaviour of the structural system. All forces and displacement were concentrated on a rather small region of the panel, which then led to local failures. The tested CLT panels behaved essentially as rigid elements. Therefore, all the dissipated energy in the testing resulted from the connections. Rinaldin et al. [7] presented experimental performance of a hold down, a bracket, and a half-lap joint connection under cyclic loading. These were tested in shear and withdrawal directions to capture the main features of the behaviour of these CLT connections. Pei et al. [8] presented the results of a series of tests that were done on CLT walls in Canada. The walls were subjected to cyclic loading and load-displacement data was recorded for the tests. The data was used for calibrating numerical model for the connections. Even though these tests have been done, all the tests were conventional CLT panels. No work on hybrid panels and connections was found in this literature review.

## 3 EXPERIMENTAL INVESTIGATION

### 3.1 HYBRID CLT PANELS AND TEST COMBINATIONS

Three- and five-layer CLT panels with hybrid layups bonded with Phenol-Resorcinol Formaldehyde adhesive (PRF) were manufactured at OSU using a 0.65m x 0.65m cold press at 0.7 MPa (100 psi). The species used in the design of experiments included one reference species, Douglas-fir (DF) (*Pseudotsuga menziesii*). Panels were built up from visually graded No. 2 and better lumbers. Low-grade lumber is represented by No. 3 and lower DF and Lodgepole Pine (LP).

The visual grade assigned was confirmed by measurement of the dynamic modulus of elasticity (DMOE) following ASTM E1876-09 [9]. The DMOE was estimated using the impulse excitation of vibration method based on the use of a Metrigard testing machine. This method was chosen since it is relatively fast, easy and non-destructive.

Pieces classified as high-grade based on visual criteria but falling below standard DMOE thresholds (i.e. 10.3 GPa for DF and 8.3 GPa for LP) were classified as low-grade material.

Table 1 lists all the combinations considered for hybrid CLT layups. The low-grade material was only used in the core of the CLT panels, while high-grade material was used in outer faces and also in the core, but only for the reference panels.

*Table 1: Species combinations for hybrid CLT layups (H: high-grade, L: low-grade, DF: Douglas-fir, LP: Lodge Pine).*

|  | Face Layer(s) |  | Core Layer(s) |  |
|---|---|---|---|---|
|  | Species | # of layers | Species | # of layers |
| **Reference** | DF(H) | 1 | DF (H) | 1 |
| **Hybrid** | DF(H) | 1 | LP (L) | 3 |
|  | DF (H) | 1 | DF(L) | 1 |

Before manufacturing, moisture content (MC) of the lumbers were checked using a WAGNER MMC 220 moisture meter. This tool uses electromagnetic waves for a non-destructive measurement of the MC. The MC should be in range of $12 \pm 3\%$ provided by PRG 320 [4]. If the MC was not in the range, the lumbers were then conditioned in a standard ASTM room (controlled 20°C and relative humidity of 65%) until the desired MC was achieved.

After grading the lumbers and controlling the MC, all the laminations were planed. The planing is essential for ensuring strong bond between the laminations by removing germ and dirt from the faces of laminations, improving the penetration of adhesive into the lumens and cell walls, and providing a uniform thickness for the laminations in the same layers. Since the panel was not edge bonded, edge planing was not needed, and only the wide faces of the lumber were planed. PRG 320 requires a maximum planing tolerance of $\pm 0.2$ mm along the width of one lamination and $\pm 0.3$ mm along the length of each single lamination. After the laminations were planed, they were cut to the desired length of 63.5 cm.

For bonding the laminations, a PRF adhesive from the products line of Hexion Inc. was chosen. This liquid adhesive is recommended for laminating softwoods and suitable for wet-use or dry-use exposure. The PRF adhesive used, Cascophen LT-75, is a liquid resin that sets through reaction with a constant proportion of a dry powdered hardener: Cascoset FM-282. The mixing proportion of 100 parts of resin and 16 parts were chosen (by weight) following recommendations by the manufacturers. The hardener was added to the resin and stirred until thoroughly dispersed, normally for about 5 minutes. Usable life (open hour) of the mixed adhesive at 16°C, 21°C, and 27°C were reported at 4.5, 2.5 and 1 hour, respectively. A spread rate of 490 g/m$^2$ (0.1 lbs/sq.ft.) was selected for spreading the adhesive in between layers. The adhesive was spread manually on the first stack using a painter metal knife and spatula. Special attention was paid to ensure a uniform adhesive spread. The next layer of laminations was then placed in the transverse direction. This process was repeated until all of the layers were stacked. The panels were then moved to the cold press and clamped under the constant pressure of 0.7 MPa (100 psi) for a minimum of 9 hours. Figure 2 illustrates the steps used for making the panels.

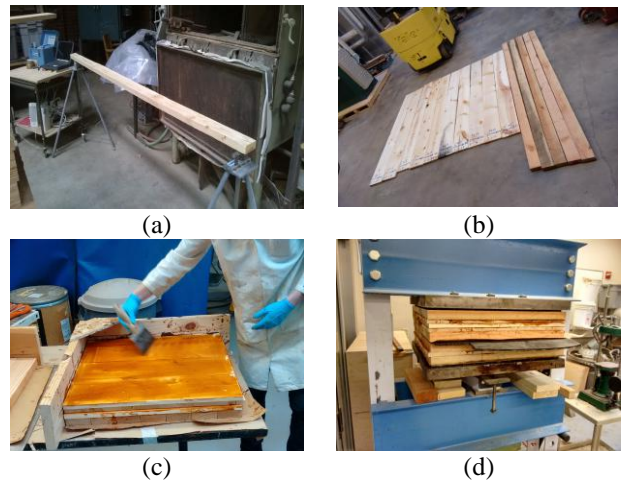

*Figure 2: Manufacturing of the panels: (a) measuring the DMOE, (b) grading the laminations (LP and DF laminations in the picture), (c) spreading the adhesive, and (d) clamping panel.*

## 4 METHODOLOGY FOR ESTIMATING CYCLIC PERFORMANCE OF CONNECTORS

A testing program to estimate the cyclic performance of connectors was performed using a Universal Testing Machine (UTM). The Simpson Strong-Tie HGA10KT bracket with four (4) SDS25112 screws on the floor side and 4 SDS25300 screws on the wall side of the connections were used for one set of tests and the ABR105 with 10 CNA4x60 and 14 CNA4x60 on the wall and floor sides, respectively. The fasteners and brackets are shown in Figure 3 and Figure 4, respectively. The connectors were connected to 0.2 m by 0.3 m cut-outs from the manufactured CLT panels. The distance of the connections from the edge of the panel for the floor side was at least 3 inches.

Table 2 shows the connector tests, panel layup, and direction of loading. Up to six (6) tests per species/grade combination and per connector type, and load direction were carried out.

### 4.1 EXPERIMENTAL SETUP AND LOADING

Figure 5 and Figure 6 illustrate the testing fixture. What is not shown in these figures are the restraints provided to the specimen to isolate forces to be developed in one direction at a time, thus isolating the response in shear and withdrawal.

For each setup, a modified CUREE [10] displacement control protocol was used. The modifications introduced were two-fold. First, for cases in which it is not realistic to model symmetric cyclic displacement, the displacement history protocol was adjusted so that only positive displacements would be developed. In addition, since the entire testing program included approximately 400 cyclic tests, in addition to the ones necessary to define the CUREE reference displacement, the loading rate for cyclic loading was adjusted so that each test could be ran in approximately five (5) minutes.

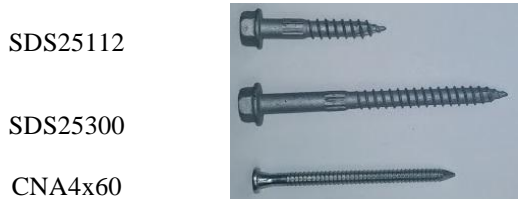

*Figure 3: The screw used in connections.*

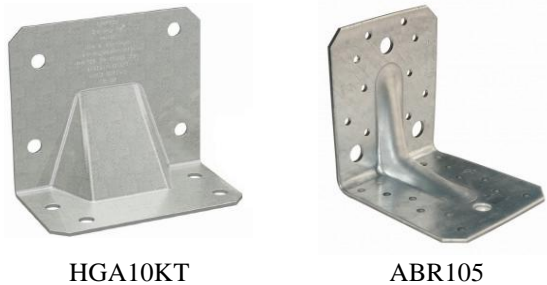

*Figure 4: The metal connector*

For tagging the tests a 9-letter long tag was used for each test. The first two characters refer to connection type (e.g. *HG* is for the HGA10KT connection and *AB* is for the ABR105 tests). The next four characters refer to panel type (*H* stands for high-grade lumber, *L* stands for low-grade lumber, *D* for Douglas-fir, and *P* for Pine). The next character shows the loading direction which *L* is for lateral (shear) and *W* for withdrawal loading direction. The remaining two digits are the test number, which ranges from 1 to 6. For instance, the *HG-HDLD-L-01* refers to the first specimen of HGA10KT brackets shear testing on hybrid panel with high grade DF laminations on the face and low-grade DF lamination in the core.

### 4.2 Instrumentation

Four channels were tracked per test, including two channels from the actuator tracking load and displacement, and two linear variable differential transformers (LVDTs) that measured uplift and rotation. The first LVDT measured the connector relative displacement was computed as the difference between the actuator displacement and the relative uplift. The latter LVDT was installed to verify that the rotations of the specimen (which were restrained from lateral movement) were negligible. Force-displacement curves and failure modes were recorded for all the tests.

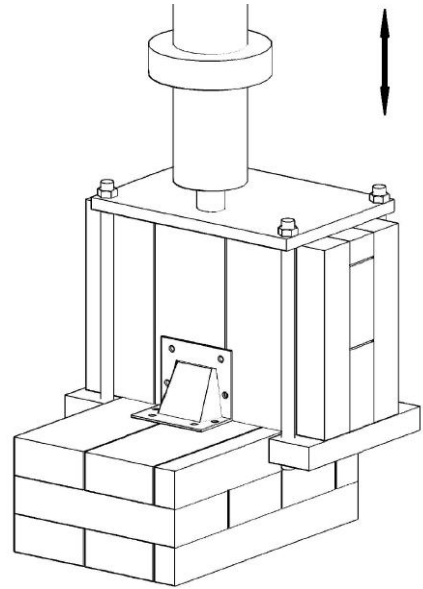

*Figure 5: Test fixture for wall-to-floor connection in withdrawal direction.*

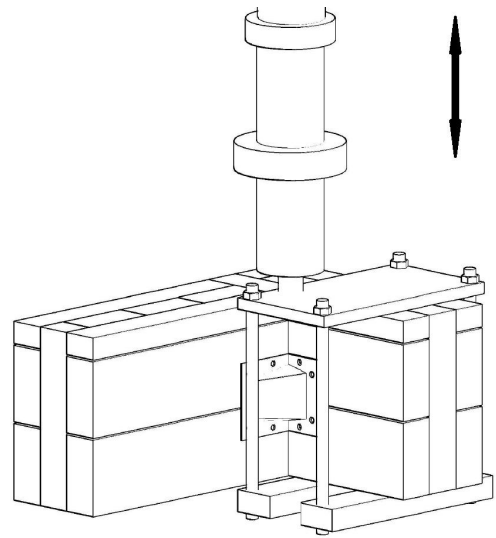

*Figure 6: Test fixture for wall-to-floor connector in shear direction.*

### 4.3 Instrumentation

Four channels were tracked per test, including two channels from the actuator tracking load and displacement, and two linear variable differential transformers (LVDTs) that measured uplift and rotation. The first LVDT measured the connector relative displacement was computed as the difference between the actuator displacement and the relative uplift. The latter LVDT was installed to verify that the rotations of the specimen (which were restrained from lateral movement) were negligible. Force-displacement curves and failure modes were recorded for all the tests.

*Table 2: Connection test configurations and test information.*

| Panel | Layup | Connectors | Load direction |
|---|---|---|---|
| Reference panels | HDHD | HGA10KT and ABR105 | shear |
| | | | withdrawal |
| Hybrid panels | HDLP | HGA10KT and ABR105 | shear |
| | | | withdrawal |
| | HDLD | HGA10KT and ABR105 | shear |
| | | | withdrawal |

### 4.4 Test Results

Force-displacement relations are the main output of the performed tests. Hysteretic plots were generated for each test. The backbone (envelope) curves were generated by connecting the maximum load points in each primary cycles. The values of maximum load and corresponding displacements are extracted from the tests results. Figure 7 presents the hysteresis plots for the tests on the HG-HDLD-L and HG-HDLD-W tests alongside with the backbone curves (shown in red). Example test results for the ABR connections are not shown in the interest of brevity.

## 5 DISCUSSION

### 5.1 Data Analysis

The values of the peak load, the displacement corresponding to the peak, and failure load, with its corresponding displacements for HGA10KT and ABR105 metal connections are obtained for each test. The failure was defined at a point where the post-peak load drops to 80% of the peak load. These values are presented in Table 3 and Table 4 for the HGA10KT and Table 5 and Table 6 for the ABR105 connectors.
In order to justify the comparison between performances of the conventional panels and the hybrid panel, a t-test used. The obtained values form the t-test are summarized in Table 7 for the HGA10KT connector and Table 8 for the ABR105 connector.
For the HGA10KT connectors, there was no significant difference between conventional and hybrid panels, both in shear and withdrawal tests (see results in Table 7). This results can be related to the penetration depth of the screws which are ~76 mm (3 inches) within the wood substrate for SDS25300, while for the SDS25112 it is only ~38mm (1.5 inches). In withdrawal tests, the failure was always imposed in the wall side of the connections, which had four SDS25112 screws. The SDS25112 screws carry load in the shear direction while the four SDS25300 screws on the floor side are carrying withdrawal loads. The withdrawal capacity of the SDS25300 screws are higher than of the shear strength of SDS25112. Since the thickness of the face layer for all the panel layups is similar, and almost equals out to ~38 mm (1.5 inches), the SDS25112 screws do not penetrate within the core layer of the panels. Therefore, the strength of all the tests ended up not being significantly different, as expected.

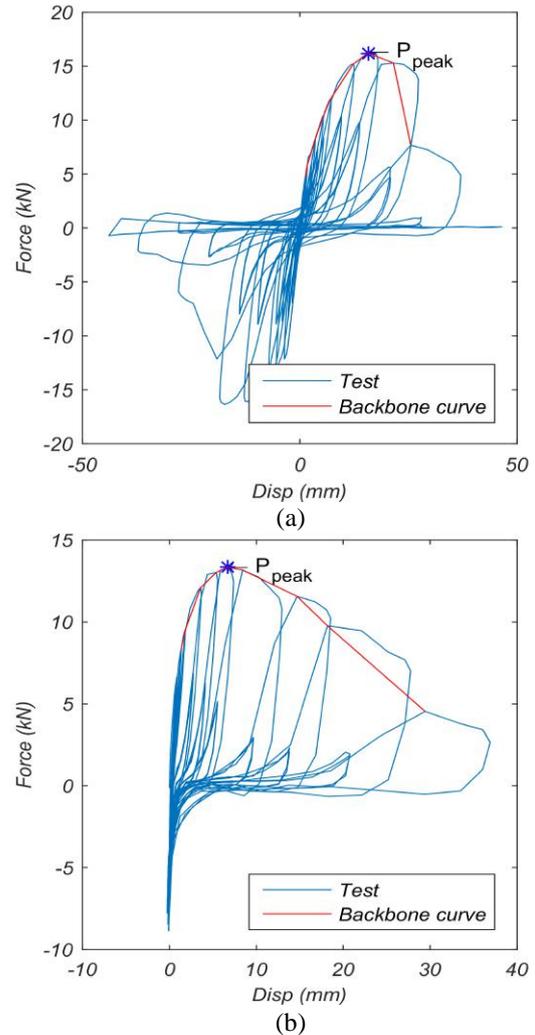

*Figure 7: The hysteresis force-displacement plots and the positive backbone curve (peak load is marked) for HGA10KT connections subjected to: (a) cyclic shear loading, and (b) withdrawal cyclic loading.*

Even though no significant difference is observed in the strength for the HGA10KT connectors, there was a statistically significant difference between the corresponding displacements at the peak load between conventional and hybrid panels. Now, the differences observed in terms of displacements need more inspection and it may be attributed to the variability of timber.
For the HGA10KT connectors, comparing HDLD vs. HDLP hybrid panels, there is no significant difference in forces or displacements between the two panel types. Considering the fact that the failure in these was observed in the SDS25112 screws, and the screws did not penetrate into the core layer, this was again expected. In the HGA10KT connections, it was observed that by having a well-designed connection, the failure will be imposed to the face layers with high-grade material.

Therefore, the low-grade material will not have a significant effect on the strength of the connection.

In the ABR connections, which were connected with CNA nails, the nails penetrated into the core layers on both floor and wall sides of the connection. By inspection of Table 8, no significant reduction was observed in shear, which may be explained on the non-even distribution of the bearing stress through the nail's shank. When the nail is loaded in shear, the bearing stress increases with increasing proximity to the nail head, and the bearing stress decreases with an increasing distance away from the nail head. In contrast, due to withdrawal loading, the whole length of the nail shank is engaged in transferring the withdrawal force in an identical form, and thus no reduction of friction between the nail's shank and the wood's substrate material is observed. Therefore the interaction between the nail and the low-grade material layer causes a reduction in the connection strength. This is clearly reflected in the statistical tests between the HDHD vs. HDLD and HDHD vs. HDLP panels.

*Table 3:* HGA10KT – values of peak load and corresponding displacement for the tests in shear loading (Loads in kN and Displacement in mm).

| Test | | Peak Load | Disp @ Peak Load | Failure Load | Disp @ Failure Load |
|---|---|---|---|---|---|
| HDHD | Mean | 14.6 | 14.0 | 11.7 | 19.9 |
|  | SD | 1.81 | 3.26 | 1.44 | 3.83 |
|  | COV | 12 % | 23 % | 12 % | 19 % |
| HDLD | Mean | 16.6 | 20.0 | 13.3 | 23.0 |
|  | SD | 1.73 | 3.41 | 1.38 | 2.02 |
|  | COV | 10 % | 17 % | 10 % | 9 % |
| HDLP | Mean | 18.2 | 20.03 | 14.6 | 21.7 |
|  | SD | 1.67 | 4. 6 | 1.34 | 6.35 |
|  | COV | 9 % | 23 % | 9.17% | 29 % |

*Table 4:* HGA10KT – values of maximum load and corresponding displacement for the tests in withdrawal loading alongside with the statistical parameters values. (Loads in kN and displacement in mm.)

| Test | | Peak Load | Disp @ Peak Load | Failure Load | Disp @ Failure Load |
|---|---|---|---|---|---|
| HDHD | Mean | 13.1 | 8.60 | 10.5 | 15.1 |
|  | SD | 1.38 | 1.21 | 1.10 | 3.02 |
|  | COV | 10 % | 14 % | 10 % | 120 % |
| HDLD | Mean | 13.5 | 10.7 | 10.8 | 17.2 |
|  | SD | 0.87 | 3.82 | 0.697 | 3.36 |
|  | COV | 6 % | 36 % | 6 % | 20 % |
| HDLP | Mean | 14.1 | 8.37 | 11.3 | 16.7 |
|  | SD | 1.61 | 3.99 | 12.85 | 2.04 |
|  | COV | 11 % | 48 % | 11 % | 12 % |

*Table 5:* ABR105 – values of maximum load and corresponding displacement for the tests in shear loading alongside with the statistical parameters values. (Loads in kN and displacement in mm.)

| Test | | Peak Load | Disp @ Peak Load | Failure Load | Disp @ Failure Load |
|---|---|---|---|---|---|
| HDHD | Mean | 20 | 30.7 | 16.0 | 40.0 |
|  | SD | 1.6 | 4.65 | 1.31 | 3.86 |
|  | COV | 8 % | 15% | 8% | 10 % |
| HDLD | Mean | 20.3 | 29.1 | 16.3 | 38.1 |
|  | SD | 1.61 | 3.90 | 1.28 | 5.52 |
|  | COV | 8 % | 13% | 8 % | 15 % |
| HDLP | Mean | 20.7 | 28.3 | 16.6 | 33.9 |
|  | SD | 1.37 | 2.88 | 1.09 | 4.55 |
|  | COV | 7 % | 10 % | 7 % | 13 % |

*Table 6:* ABR105 – values of maximum load and corresponding displacement for the tests in withdrawal loading alongside with the statistical parameters values. (Loads in kN and displacement in mm.)

| Test | | Peak Load | Disp @ Peak Load | Failure Load | Disp @ Failure Load |
|---|---|---|---|---|---|
| HDHD | Mean | 21.7 | 10.2 | 17.3 | 15.9 |
|  | SD | 0.99 | 1.38 | 0.80 | 0.99 |
|  | COV | 5% | 14 % | 5% | 6 % |
| HDLD | Mean | 19.2 | 9 | 15.3 | 27.8 |
|  | SD | 1.49 | 1.41 | 1.12 | 8.04 |
|  | COV | 8 % | 16 | 8 % | 29% |
| HDLP | Mean | 19.4 | 8.61 | 15.5 | 16.0 |
|  | SD | 1.40 | 0.85 | 1.18 | 4.04 |
|  | COV | 8% | 10% | 8% | 25 % |

*Table 7:* HGA10KT – Result of t-test on the connections for peak loads and the corresponding displacements (Loads in kN and Disp in mm; Yes and No shows if the t-test shows any significant difference between the two tests or not)

| T-test | | Shear | | Withdrawal | |
|---|---|---|---|---|---|
|  |  | Disp | Load | Disp | Load |
| HDHD vs. HDLD | Difference | **Yes** | No | No | No |
|  | p-value | 0.0221 | 0.0789 | 0.2137 | 0.562 |
| HDHD vs. HDLP | Difference | **Yes** | **Yes** | No | No |
|  | p-value | 0.0344 | 0.005 | 0.891 | 0.275 |
| HDLD vs. HDLP | Difference | No | No | No | No |
|  | p-value | 0.896 | 0.1342 | 0.979 | 0.928 |

*Table 8: ABR105 – Result of t-test on the connections for peak loads and the corresponding displacements (Loads in kN and Disp in mm; Yes and No shows if the t-test shows any significant difference between the two tests or not.)*

| T-test | | Shear | | Withdrawal | |
|---|---|---|---|---|---|
| | | Disp | Load | Disp | Load |
| HDHD vs. HDLD | Difference | No | No | No | **Yes** |
| | p-value | 0.5330 | 0.7528 | 0.2652 | 0.0360 |
| HDHD vs. HDLP | Difference | No | No | No | **Yes** |
| | p-value | 0.3433 | 0.4613 | 0.0850 | 0.0488 |
| HDLD vs. HDLP | Difference | No | No | No | No |
| | p-value | 0.7133 | 0.6717 | 0.6027 | 0.8250 |

### 5.2 The Failure Modes

The failure modes are described for the HGA10KT connections only. The reader can contact the authors for information on the ABR105.

Several modes of failure were observed during the tests. In shear tests, the failure was observed in the wall side of the connection, which had four SDS25300 screws. The failure was due to crushing of the fibres in wood substrate material. Figure 8 represents the observed mode of failure. The failure in the screws resembles mode $I_m$ of the European yield model from the NDS. The screws almost always were left intact. The connection was slightly damaged and bent in the direction of the loading.

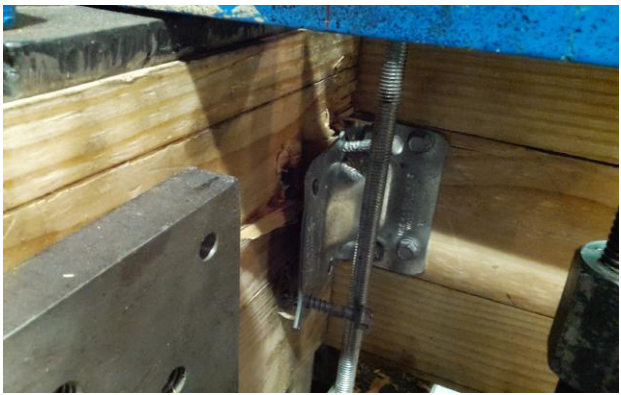

*Figure 8: HGA10KT – Observed mode of failure in the shear tests at SDS25300 screws wood substrate material on the wall side of the panel*

In withdrawal, the failures were always on the wall side of the connection at the SDS25112 screws in the wood substrate material also resemble mode $I_m$ of the European yield model from the NDS. The damage was similar to the one observed in shear loading, but the only difference was that the damage was seen here along the direction of grain, while for the shear the failure was perpendicular to the grain direction. No damage was visible on the floor side of the connections. Figure 9 shows this mode of failure.

In some cases, the failures observed were result of splitting of a single lamination piece on the wall side (Figure 10). In this case, one row of screws were spaced in fraction of an inch from the gap between the laminations.

## 6 Conclusions

The results of a cyclic testing program on select connector system (Simpson Strong-tie HGA10KT and ABR105) on hybrid CLT layups for wall-to-floor connections are presented in this paper. The performance of connectors in two perpendicular directions under modified CUREE cyclic loading protocol were presented. The tests showed that difference in the strength of the connection systems, on hybrid layup was not statistically significant from the conventional panels for the HGA10KT. However, for the ABR105 withdrawal tests, strength results can be considered statistically significantly different.

It is clear for these tests that the detailing of the connectors are important to the performance of the connectors. With an appropriate selection of the connection system, and if the failure is imposed to occur on the face layer of the hybrid panel, the corresponding low-grade core layers will have negligible effects on the strength of the panels.

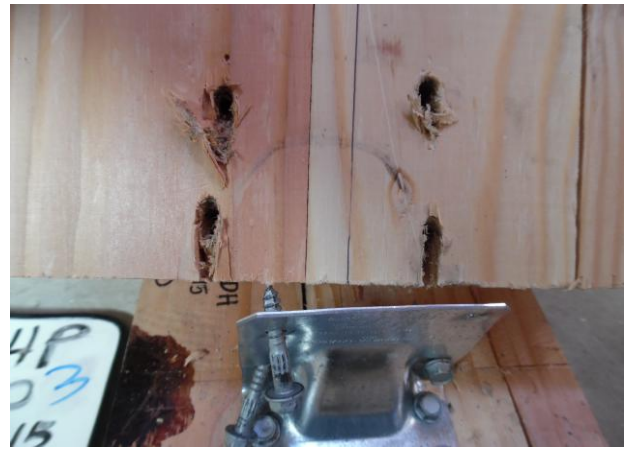

*Figure 9: HGA10KT – The observed mode of failure in the withdrawal test.*

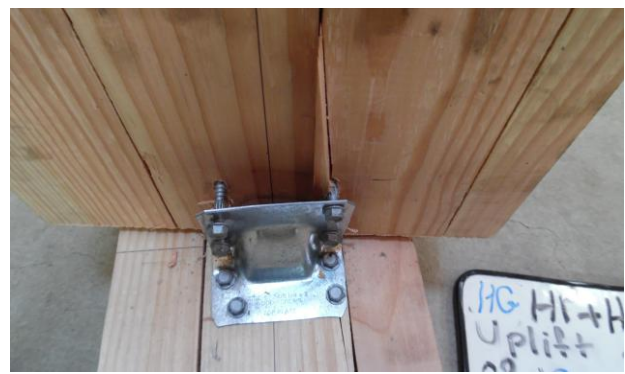

*Figure 10: HGA10KT – Observed mode of failure in withdrawal tests. In this case, splitting in a single lamination is observed due to the small distance of the screws to the gap between the laminations.*